\documentclass[12pt,3p]{article}
\usepackage[utf8]{inputenc}
\usepackage{graphicx} 
\usepackage{geometry}
\usepackage{appendix}
\usepackage{chemarrow}
\usepackage{extarrows}
\usepackage{mathtools}
\usepackage{tikz}
\usetikzlibrary{arrows,automata}
\usepackage{amsmath}
\usepackage{amssymb}
\usepackage{natbib}
\usepackage{setspace}

\begin{document}

\title{DNA methylation heterogeneity induced by collaborations between enhancers} 

\author{YUSONG YE$^1$,\quad
ZHUOQIN YANG$^1$,\quad and
JINZHI LEI$^{2}$\footnote{Corresponding author, Email: jzlei@tsinghua.edu.cn}
}
\date{}
\maketitle

\begin{center}
\begin{enumerate}
\item[$^1$]{School of Mathematics and Systems Science and LMIB, Beihang University, Beijing 100191, China}      
\item[$^2$]{Zhou Pei-Yuan Center for Applied Mathematics, MOE Key Laboratory of Bioinformatics, Tsinghua University, Beijing 100084, China}
\end{enumerate}
\end{center}

\vspace{0.5cm}

\begin{spacing}{2.0}

\begin{abstract}
During mammalian embryo development, reprogramming of DNA methylation plays important roles in the erasure of parental epigenetic memory and the establishment of na\"{i}ve pluripogent cells. Multiple enzymes that regulate the processes of methylation and demethylation work together to shape the pattern of genome-scale DNA methylation and guid the process of cell differentiation.  Recent availability of methylome information from single-cell whole genome bisulfite sequencing (scBS-seq) provides an opportunity to study DNA methylation dynamics in the whole genome in individual cells, which reveal the heterogeneous methylation distributions of enhancers in embryo stem cells (ESCs). In this study, we developed a computational model of enhancer methylation inheritance to study the dynamics of genome-scale DNA methylation reprogramming during exit from pluripotency. The model enables us to track genome-scale DNA methylation reprogramming at single-cell level during the embryo development process, and reproduce the DNA methylation heterogeneity reported by scBS-seq. Model simulations show that DNA methylation heterogeneity is an intrinsic property driven by cell division along the development process, and the collaboration between neighboring enhancers is required for heterogeneous methylation.  Our study suggest that the mechanism of genome-scale oscillation proposed by Rulands et al. (2018) might not necessary to the DNA methylation during exit from pluripotency.
\end{abstract}


\section{Introduction}
In mammalian development, reprogramming of DNA methylation (5-methylcytosine) patterns play a crucial role in defining cell fate. Upon fertilization, DNA methylation marks represent an epigenetic barriers that restrict mammalian development, and hence need to be restored and subsequently rebuilt with the commitment to  particular cell fates\citep{seisenberger2013reprogramming}. The segregation of cell lineages give rise to different somatic tissues associated with tissue-specific DNA methylation patterns \citep{styblo2000comparative, Greenberg:2019hw, Hon:2013jr}. The genome-wide DNA methylation reprogramming events coincide with the changes in concentrations of DNA methyltranferases (DNMTs) and  the enzymes that initiate the removal of DNA methylation (ten-eleven-translocation family proteins, TETs)\citep{seisenberger2013reprogramming,Greenberg:2019hw}.  The recent maturation of single-cell sequencing technologies has enable us to observe a variety of sequencing information at individual cells level, such as the genome, transcriptome, and epigenome \citep{Stuart:2019iv}. It becomes a challenge issue in computational biological to develop single-cell based computational model that can help us to better understand the process of DNA methylation pattern formation as well as cell fate decision during early embryo development.

The availability of methyleome information from single-cell whole genome bisulfite sequencing (scBS-seq) provides an opportunity to study DNA methylation patterns in the whole genome in individual cells\citep{Farlik:2015bw,rulands2018genome,Smallwood:2014kn}.  A recent study applied scBS-seq to embryo stem cells (ESCs) cultured under na\"{i}ve (two chemical inhibitors (``2i'') of MEK1/2 and GSK3$\alpha$/$\beta$) and primed (``serum'') conditions to explore DNA methylation dynamics in cells undergoing a biological transition\citep{rulands2018genome}, primed ESCs had increased variance at several genomic annotations associated with active enhancer elements, including H3K4me1 and H3K27ac sites and low methylated regions (LMRs). Analysis of scBS-seq data shown that individual primed ESCs have average DNA methylation levels varying between 17\% and 86\% at enhancers, while na\"{i}ve ESCs showed minimal cell-to-cell variability, and DNA methylation heterogeneity was resolved upon differentiation to embryoid bodies\citep{rulands2018genome}. In \cite{rulands2018genome}, it was proposed that the DNA methylation heterogeneity is associated with coherent, genome-scale oscillations in DNA methylation, and amplitude is dependent on the CpG density.  Moreover, a mathematical model of delay differential equation with autocatalytic \textit{de novo} methylation was proposed to show that global oscillations may emerge from the biochemistry of methylation turnover due to Hopf bifurcation with increasing values of the time delay, and a Kuramoto model was applied to describe the global heterogeneous coupling of CpGs via DNMT3a/b binding. Nevertheless,  genome-scale oscillations in DNA methylation is a very strong assumption, which may imply global oscillations in transcriptions of most genes, and is not supported by the experimental data  of DNA methylation dynamics during transition from na\"{i}ve to primed pluripotency \textit{in vitro} (detailed below). Hence, we asked how the transitions of DNA methylation heterogeneity should be explained through a simple mechanism? 

DNA methylation and chromatin dynamics have been modeled quantitatively in various genomic contexts of biological significance\citep{Berry:2017km,haerter2013collaboration,Huang:2017jr,Sneppen:2011db,song2017collaborations}. The collaboration between neighboring CpGs was highlight in recent studies\citep{haerter2013collaboration,song2017collaborations}, which play essential roles in the formation of global patterns and the genome-scale transitions of DNA methylation. The collaboration may directly come from the binding of DNMT3a/b to neighboring CpGs\citep{rulands2018genome}, or indirectly through the interaction with methylations in the histones H3K9\citep{Lehnertz:2003eb} and H3K36\citep{Weinberg:2019gv}. In additional to the \textit{de novo} methylation and demethylation, dilution and maintenance of of methylated marks during DNA replication may also contribute to the cell-to-cell variance of DNA methylation.  

Here, motivated by the scBS-seq data of heterogeneous methylation at genomic annotations associated with active enhancer elements, we developed a  model of DNA methylation, considering the stochastic dynamics methylation levels of enhancers over cell divisions and the collaboration between neighboring enhancers, to investigate the transition of DNA methylation from  na\"{i}ve to primed ESCs. The model focus at the random inheritance of DNA methylations during cell cycling, and can automatically reproduce the DNA methylation heterogeneity on enhancers during embryonic development. Our results suggest that the mechanism of genome-scale oscillation proposed by \cite{rulands2018genome} may not required for the observed heterogeneity during exit from pluripotency. 

\section{Results}

\subsection{Transition of DNA methylation patterns from na\"{i}ve to primed ESCs}

\begin{figure}[htbp]
	\centering 
	\includegraphics[width=14cm]{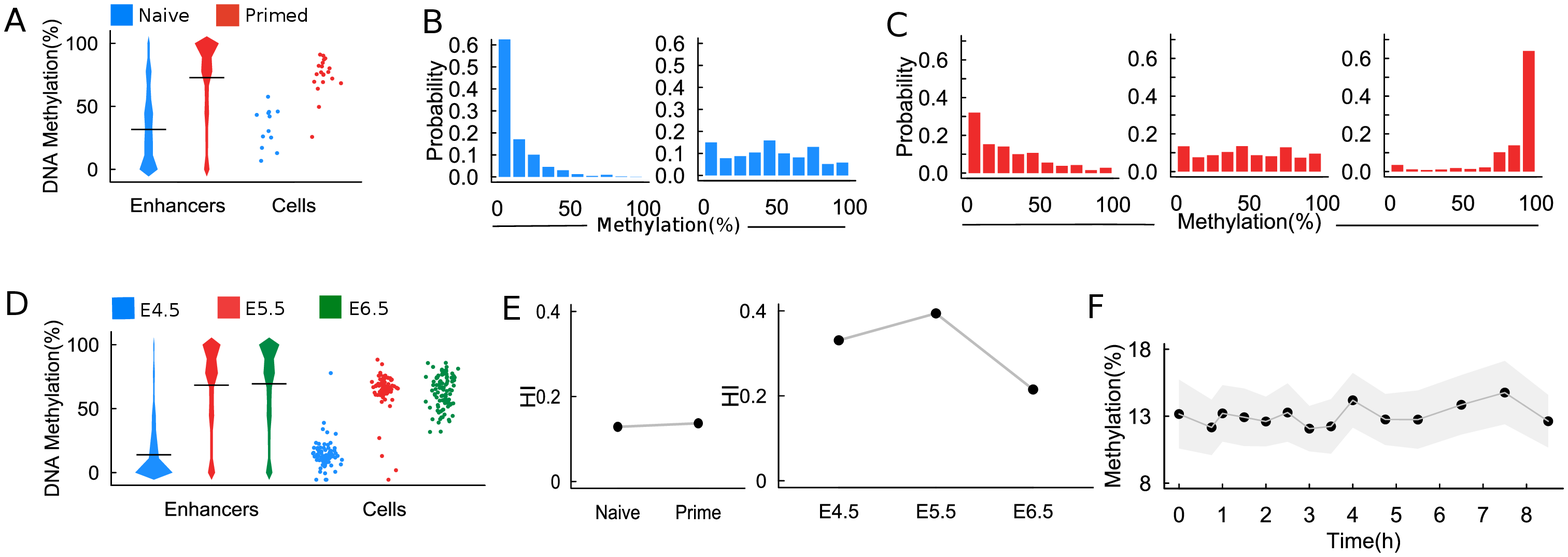}  
\caption{\textbf{DNA methylation transition in embryo cells.} (A). DNA methylation variance in na\"{i}ve and primed ESCs compared  the enhancer elements defined using published H3K4me1 ChIP-seq data\citep{creyghton2010histone}, and the average methylation levels in individual cells. (B). Distributions of enhancer methylation levels in two cells from the na\"{i}ve condition. (C). Distributions of enhancer methylation levels in three cells from the primed condition. (D). DNA methylation variance of epiblast cells at E4.5, E5.5, and E6.5 of mice embryo, and the average methylation levels in individual cells. (E). Heterogeneity indexes of cells under different conditions. (F). Dynamics of average DNA methylation from $0$ to $8$ h in the ``2i release'' experiments. Here the methylation levels were calculated from the average over the enhancers at ch1.  All data were obtained from \cite{rulands2018genome}. }
\label{fig:1}
\end{figure}

We analyzed the scBS-seq data separately for ESCs cultured under naive (‘‘2i’’) and primed (‘‘serum’’) conditions\citep{rulands2018genome}. Similar to the analysis in \citep{rulands2018genome}, taking published H3K4me1 chromatin immunoprecipitation sequencing (ChIP-seq) data form primed ESCs\citep{creyghton2010histone} as a definition of enhancer elements, the methylation levels of enhancers in primed ESCs increase comparing with na\"{i}ve ESCs (Fig. \ref{fig:1}A). Here, the methylation level of an enhancer is defined as the average level of all CpG sites contained in the enhancer. For each CpG site, we assigned a value $0$ for unmethylated, $0.5$ for half-methylated, and $1$ for full methylated, hence the methylation level of an enhancer takes  value from the interval $[0, 1]$ (or from $0\%$ to $100\%$ methylated).  Moreover, we calculated the distribution of methylation levels of all enhancers in individual cells. Primed ESCs shown higher cell-to-cell variability at the distribution patterns than the na\"{i}ve ESCs (Fig. \ref{fig:1}B \& C). We also analyzed the parallel scM\&T sequencing of \textit{in vivo} epiblast cells at E4.5, E5.5, and E6.5\citep{rulands2018genome}, which shown an increase in the methylation level in enhancers from E4.5 to E5.5 (Fig. \ref{fig:1}D). We note that there are a few cells shown low methylation levels at E5.5, however all cells have high methylation at E6.5, which suggest a transition dynamics of methylation levels  (Fig. \ref{fig:1}D). 
 
To quantify the heterogeneity of DNA methylations among different cells, we proposed a definition of heterogeneity index based on the methylation levels of enhancers in each cell.  Assuming that there are $n$ cells, and $p_i$ ($1\leq i \leq n$) the distribution of all enhancer methylation levels of the $i$'th cell, we defined the heterogeneity index ($H$) as the average of Kullback-Leibler divergence between any two cells. Mathematically, the heterogeneity index is formulated as
\begin{equation}
\label{eq:HI}
H = \dfrac{1}{n(n-1)}\sum^n_{i,j=1}KL(p_i || p_j),
\end{equation}
where $KL(p_i||p_j)$ means the Kullback-Leibler divergence between the distributions of enhancer methylation levels for the two cells $i$ and $j$,
$$KL(p_i ||p_j) = \int_{0}^{1} p_i(x) \log \dfrac{p_i(x)}{p_j(x)} dx.$$

We calculated the heterogeneity index based on the above data from na\"{i}ve and primed ESCs, and the cells at E4.5, E5.5, and E6.5 mice embryo. The is no significant changes in the heterogeneity of na\"{i}ve ESCs in comparing with the primed ESCs (Fig. \ref{fig:1}E). For the mice embryo cells, the heterogeneity index increases from E4.5 to E5.5, and decreases from E6.6 to E6.5 (Fig. \ref{fig:1}E). 

To validate the genome-scale DNA methylation oscillations in \cite{rulands2018genome}, we analyzed the same data from an \textit{in vitro} ``2i release'' model in which cells were transferred from na\"{i}ve 2i to primed serum culture conditions. The average methylation levels of enhancers from the first chromosome (ch1) were calculated to obtain the dynamics from $0$ to $8$ h after 2i release.  The results do not show significant oscillations in the methylation level (Fig. \ref{fig:1}F). This findings suggest that the assumption of genome-scale DNA methylation oscillations might not necessary to explain the transition of methylation and heterogeneity from na\"{i}ve to primed ESCs. 

\subsection{Stochastic dynamics of enhancer methylation levels}

The dynamics of DNA methylation/demethylation have been modeled quantitatively with exquisite details at biochemistry of single CpG sites\citep{haerter2013collaboration,song2017collaborations,Lei:2018ev,Sontag:2006iw}.  The methylation state of a single CpG site is often random  due to the stochastic biochemical reactions. Nevertheless, the average methylation level of CpG sites associated with a genomic segment is more predictable. During embryo development, the most significant changes in DNA methylation occur during DNA replication when the 5-methylcytosine marks are dilute to two daughter strains and are restored through enzymes DNMT1 and nuclear protein 95 (NP95 or UHRF1). Correlating global DNA methylation with replication timing repli-seq data shown that late-replicating regions did not have lower DNA methylation than early-replicating regions\citep{rulands2018genome}. Thus, while we omit the details dynamics between DNA replications, we can represent the methylation level of an enhancer by the average methylation level at late-replication stage of each cell cycle. 

\subsubsection{Formulation}
To consider the dynamics of enhancer methylation levels, assuming that there are $N$ enhancers in a chromatin, and letting $\beta_i^t$ the methylation level of the $i$'th enhancer at cycle $t$ ($0\leq \beta_i^t \leq 1$), we only need to formulate the dynamics of the states
$$\vec{\beta}^t = (\beta_1^t, \beta_2^t, \cdots, \beta_N^t)$$
over cell cycles $t$. During cell cycling, the methylation states update as a consequence of the regulations through enzymes DNMT1, DNMT3a/b and TETs, which leads to the following iteration
\begin{equation}
\label{eq:1}
T: \left( \beta^t_1, \beta^t_2,\beta^t_3, ...,   \beta^t_N \right) \xrightarrow{\mbox{cell cycle}}\left( \beta^{t+1}_1, \beta^{t+1}_2,\beta^{t+1}_3, ...,  \beta^{t+1}_N \right).
\end{equation}
Hence, while we omit the biochemistry details, the stochastic dynamics of enhancer methylation levels can be formulated as an iteration
\begin{equation}
\label{eq:m}
\vec{\beta}^{t+1} = T(\vec{\beta}^t)
\end{equation}
for each cell cycle. Here, $\vec{\beta}^{t}$ represents the methylation state of a cell before cell division, and $\vec{\beta}^{t+1}$ is the state of one daughter cell after cell division. 

We note that the iteration Eq. \eqref{eq:m} is usually a random map. Given the state of a mother cells, the methylation state of the daughter cell is a random valuable whose probability density is dependent on the state of the mother cell.  Hence, to formulate the iteration map, we need to write down the conditional probability density function
\begin{equation}
\mathrm{Prob}(\vec{\beta}^{t+1} = \vec{x}| \vec{\beta}^t),
\end{equation}
the probability of $\vec{\beta}^{t+1}$ given the state of the mother cell $\vec{\beta}^t$. While the probability of each enhancer, given the state of the mother cell, is independent to each other, we have
\begin{equation}
\label{eq:4}
\mathrm{Prob}(\vec{\beta}^{t+1} = \vec{x}| \vec{\beta}^t) = \prod_{i=1}^N \mathrm{Prob}(\beta_i^{t+1} = x_i | \vec{\beta}^t).
\end{equation}

The probability density $\mathrm{Prob}(\beta_i^{t+1} = x_i | \vec{\beta}^t)$ is usually not known and may depend on the biochemistry details of methylation/demethylation. Nevertheless, it is possible to write down the phenomena formulation if we overlook the detail process. If there are $m_i$ CpGs in the $i$'th enhancer, each CpG has a probability $p_i^t$ to be methylated (here the superscript $t$ specified the dependence with the time $t$), and $1-p_i^t$ to be unmethylated after cell division (here we omitted the state of half-methylation), the probability to have $k_i$ methylated CpGs is given by a binomial distribution  $C_{m_i}^{k_i}(p_i^t)^{k_i} (1-p_i^t)^{m_i-k_i}$ with the parameters $p_i^t$. To extend the probability to a more general situation, the binomial distribution can be generalized to a beta-binomial distribution through two shape parameters $a_i^t$, $b_i^t$. Moreover, while we are only interested at the probability of the methylation level $x_i = k_i/m_i$, we replaced the beta-binomial distribution with the beta distribution, and hence
\begin{equation}
\label{eq:5}
\mathrm{Prob}(\beta_i^{t+1} = x_i | \vec{\beta}^t) = \dfrac{x^{a_i^t-1} (1-x_i)^{b_i^t-1}}{B(a_i^t, b_i^t)},\quad B(a, b) = \dfrac{\Gamma(a)\Gamma(b)}{\Gamma(a+b)},
\end{equation}
where $\Gamma(\cdot)$ is the gamma function,  $a_i^t$ and $b_i^t$ are shape parameters depending on $\vec{\beta}^t$. Eqs. \ref{eq:4}--\ref{eq:5} together define the conditional probability $\mathrm{Prof}(\vec{\beta}^{t+1} = \vec{x} | \vec{\beta})$. The beta distribution is one of few common ``named'' distributions that give probability $1$ to a finite interval. As the shape parameters $a$ and $b$ vary, the beta distribution can take different shapes, include strictly decreasing ($a \leq 1, b > 1$), strictly increasing ($a > 1, b \leq 1$), U-shaped ($a < 1, b < 1$), or unimodal ($a> 1, b>1$).  Thus, the dependences of parameters $a_i^t, b_i^t$ on the state $\vec{\beta}^t$ are important to define the shape of the distribution function.

Now, to define the dependences $a_i^t(\vec{\beta}^t)$ and $b_i^t(\vec{\beta}^t)$, assuming that there are functions $\phi_i^t(\vec{\beta}^t)$ and $\eta_i^t$ so that average of $\beta_i^{t+1}$, given $\vec{\beta}^t$, is
$$\left.\langle \beta_i^{t+1}\rangle\right|_{\vec{\beta}^t} = \phi_i^t(\vec{\beta}^t),$$
and the variance 
$$\left.\mathrm{Var}(\beta_i^{t+1})\right|_{\vec{\beta}^t} = \dfrac{1}{ 1 + \eta_i^t} \left(1-\phi_i^t(\vec{\beta}^t)\right) \phi_i^t(\vec{\beta}^t),$$
than\footnote{Here we note that 
$$\left.\langle \beta_i^{t+1}\rangle\right|_{\vec{\beta}^t} = \dfrac{a_i^t}{a_i^t + b_i^t},$$
and
$$\left.\mathrm{Var}(\beta_i^{t+1})\right|_{\vec{\beta}^t} = \dfrac{a_i^t\, b_i^t}{ (a_i^t + b_i^t) (1 + a_i^t + b_i^t)}.$$
}
\begin{equation}
\label{eq:ab}
a_i = \eta_i^t \phi_i^t(\vec{\beta}^t),\quad b_i^t = \eta_i^t \left(1 - \phi_i^t(\vec{\beta}^t)\right).
\end{equation}
Hence, we only need to identify the functions $\phi_i(\vec{\beta}^t)$ and $\eta_i^t$, and the parameters $a_i^t$ and $b_i^t$ can be defined accordingly. Specifically, we take $\eta_i^t$ as the number of CpGs in the $i$'th enhancer, \textit{i.e.},
\begin{equation}
\eta_i^t = m_i.
\end{equation} 
This simple assumption means the inverse proportion of the variance of enhancer methylation levels with the number of CpGs.   

To define the function $\phi_i^t$, we assumed the methylation level of daughter cell depends on that of the mother cell through three components: basal methylation level, autocatalytic effect, and collaboration between enhancers. Hence, the function $\phi_i^t$ was formulated as
\begin{equation}
\label{eq:6}
\phi_i^t(\vec{\beta}^t)=H_{[0,1]}(z),\quad z=\underbrace{\mu_0}_{\mathrm{basal}}+\underbrace{\mu_1\dfrac{(\beta^t_i)^n}{(\beta^t_i)^n+v}}_{\mathrm{autocatalysis}}+\underbrace{\dfrac{\alpha}{2L+1}\sum\limits_{|j-i|\leq L}({\beta^{t}_j }-\beta^t_i)}_{\mathrm{collaborations}},
\end{equation}
where
\begin{equation}
H_{[0,1]}(z) = \left\{
\begin{array}{ll}
0,&\quad z<0,\\
z,&\quad 0\leq z\leq 1\\
1,&\quad z>1.
\end{array}\right.
\end{equation}
Here $\mu_0, \mu_1, n, v, \alpha, L$ are parameters, with $\mu_0$ the basal level,  $\mu_1$ the coefficient for autocatalysis, $n$ the Hill coefficient, $v$ the parameter for the autocatalytic efficiency, $\alpha$ the coefficient for long distance collaboration, and $L$ defines the range of collaboration between enhancers. The autocatalytic efficiency $v$ usually depends on the activity of \textit{de novo} methylation/demethylation regulated by DNMT3a/b and TETs, and hence can change with time $t$ during embryo development. Here, we always have $0\leq \phi_i^t \leq1$ due to the function $H_{[0,1]}(z)$.

The third term in Eq. \ref{eq:6} shows the collaboration between neighboring enhancers. Here, we assumed the coherent collective behaviors of DNA methylation/demethylation when the average coupling through enzymes binding is sufficiently strong so that the nearby enhancers tend to the same trends of either methylation or demethylation. The similar mechanism was introduced previously to reproduce the long distance correlation of DNA methylation between CpGs\citep{song2017collaborations}. Here the collaboration effect is limited by cooperative range $L$ and the coefficient $\alpha$.

\subsubsection{Numerical scheme and parameters}
To model the methylation dynamics following developmental process with the above iteration Eq. \ref{eq:1}, we first initialized the methylation level of $N$ enhancers $\vec{\beta}^0 = (\beta_1^0, \beta_2^0, \cdots, \beta_N^0)$ (here $t=0$). Here, each enhancer $i$ associates with an integer $m_i$ for the number of CpGs. Next, at each step, we calculated the shape parameters $a_i^t$ and $b_i^t$ using Eqs. \ref{eq:ab}--\ref{eq:6}. Finally, for each enhancer $i$, generated a random number in according to the beta distribution Eq. \ref{eq:5}, and set $t=t+1$. We repeated the above scheme to generate the dynamics of methylations in each enhancer following multiple cell cycles. 

\begin{figure}[htbp]
	\centering 
	\includegraphics[width=14cm]{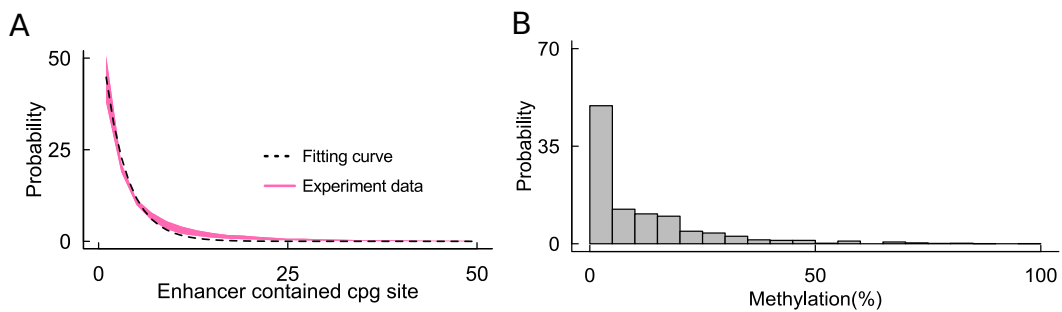}  
\caption{{\bf Number of CpG sites and the initial methylation level.}
(A). Distribution of number of CpG sites in enhancers and the fitting curve (Eq. \eqref{eq:fit}). (B). Distributions of methylation at a na\"{i}ve ESC.}
\label{fig:7}
\end{figure}

In model simulations, we took $\mu_0 = 0.01, \mu_1 = 0.9, n = 3, \alpha = 0.5, L=50$, and varied the autocatalysis ability $v$ to represent different culture conditions ($v = 0.1, 0.04, 0.01$ for situations of high, mediate, and low level DNA methylation, respectively). Moreover, to mimic an enhancers in a chromosome, we performed simulations with $N=1000$ enhancers, the CpG numbers for each enhancer were taken following an exponential distribution 
\begin{equation}
\label{eq:fit}
\mathrm{Prob}(m_i=m)=\frac{e^{(-m/3)}}{1.6}. 
\end{equation}
in according to the statistics from mouse genome(Fig. \ref{fig:7}A). In simulations, we can refer the distribution of methylation levels from a na\"{i}ve as the initial condition(Fig. \ref{fig:7}B).

\subsection{Transition of DNA methylation heterogeneity during the development process}

\begin{figure}[htbp]
	\centering 
	\includegraphics[width=14cm]{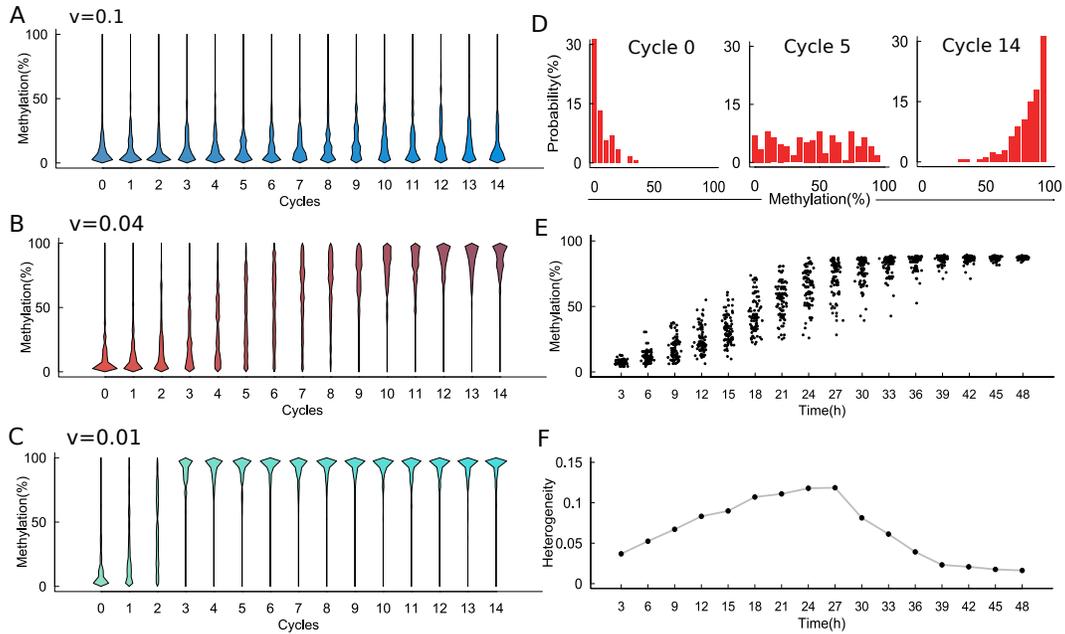}  
\caption{\textbf{DNA methylation heterogeneity from model simulation.}
(A-C).  Distributions (violin plots) of methylation levels in enhancers obtained from model simulation with $v=0.1$ (A), $v=0.04$ (B), and $v=0.01$ (C).  (D). Histogram methylation levels in enhancers for cycles 0, 5, and 14. (E). Transition of DNA methylation from low to high level. Each dot represent the average methylation level in a cell. (F). Evolution of the heterogeneity index (HI). The HI were calculated from the cells shown by (E). Here $v=0.04$ in (E)-(F).}
\label{fig:3}
\end{figure}

To verify the proposed model, we varied the parameter $v$ ($v = 0.1, 0.04, 0.01$) to mimic different conditions. For each value $v$,  we initialized a cell with an initial state of a na\"{i}ve cell and ran the model simulation for 15 cell cycles in order to mimic the transition from na\"{i}ve to primed condition. The simulated distribution of enhancers methylation levels  at each cell cycle were calculated, and are shown by Fig. \ref{fig:3}A-C. The enhancers methylation distributions depend on the parameter $v$: when $v=0.1$, the enhancers were homogeneous with low level methylation; when $v=0.04$, the cells shown obvious methylation heterogeneity, the methylation levels transfer from low to high along with cell cycling; when $v=0.01$, most enhancers change to high level methylations in a few cycles. When $v=0.04$, we have low, mediate, or high methylation levels in the enhancers at different cycles (Fig. \ref{fig:3}D). These results are in agree with experimental observations of ESCs cultured under serum condition, and hence the model can be used to mimic the DNA methylation heterogeneity of ESCs during exist from pluripotency. 

To further examine the transition dynamics of methylation heterogeneity in ESCs, we set $v=0.04$ and initialized a population of cells according to the methylation distribution at Fig. \ref{fig:7}B. Next, we performed the simulation scheme to mimic a development process of 48 h. In simulations, each cell divides with a probability of $0.3$ per h, and collected a subpopulation of cells to calculate the average methylation level and the heterogeneity index every 3 h.  Simulations shown that the average methylation level increased from $0$ to $48$ h, but shown obvious diversity from $12$ to $30$ h (Fig. \ref{fig:3}E). The heterogeneity index shown non-monotonous with the development process, firstly increase from $0$ to $27$ h to reach a high level of $0.12$, and then decrease to a low level heterogeneity at 48h (Fig. \ref{fig:3}F). In embryo development, DNA methylation heterogeneity in ESCs increased from na\"{i}ve to primed, and the heterogeneity was resolved upon differentiation to embryoid bodies \citep{rulands2018genome}. These results shown similar dynamics in both experiments and our model simulations. 

\subsection{Collaboration and methylation heterogeneity}

\begin{figure}[htbp]
	\centering 
	\includegraphics[width=12cm]{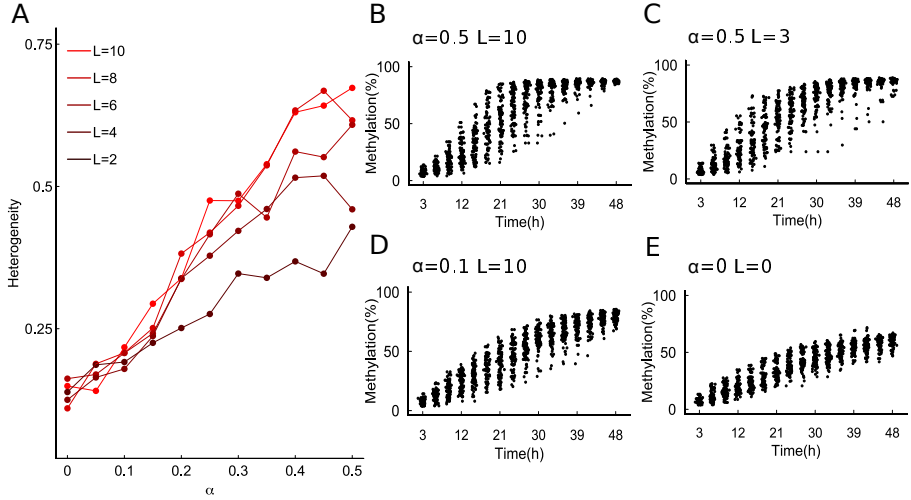}  
\caption{{\bf Collaboration and methylation heterogeneity.}
(A). Heterogeneity index with different values of the parameter $\alpha$ and $L$. (B-D) Transition dynamics of the average methylation level with varied parameter values of $\alpha$ and $L$.}
\label{fig:4}
\end{figure}

The proposed model includes collaboration between neighboring enhancers so that there are coherent collective behaviors during enzyme binding. To investigate the effects of neighboring collaboration, we varied the parameters $\alpha$ and $L$ and examined the changes in methylation heterogeneity. Here, the parameter $\alpha$ measures the collective strength, and $L$ gives the regions of neighboring collaboration. Simulations shown that the heterogeneity increased with $\alpha$ for different values of $L$, and increased with $L$ for a large value $\alpha$ (Fig. \ref{fig:4}A).  

To further examine how collaborations may affect the DNA methylation dynamics, we varied the parameters $\alpha$ and $L$ and calculated the evolution dynamics over a period of $48$ h. When $\alpha = 0.5$ and $L=10$, the cells shown highly heterogeneous during the intermediate transition region (Fig. \ref{fig:4}B). When either $L$ or $\alpha$ decreases, the cells shown less heterogeneity (Fig. \ref{fig:4}C-D). In particular, when $\alpha=0$, which represents the situation without collaboration, all cells shown similar DNA methylation dynamics during simulation from $0$ to $48$ h, with the average methylation increases from a low level to an intermediate level of $50$\% (Fig. \ref{fig:4}E). These results suggest that the collaboration between neighboring enhancers is required to produce DNA methylation heterogeneity.  

\section{Discussion}

Reprogramming of DNA methylation plays important roles in mammalian early embryo development. ESCs show heterogeneous methylation distributions under primed conditions. To understand the mechanism of methylation heterogeneity, previous studies suggest a mechanism of genome-scale oscillations in DNA methylation. Nevertheless, experiment data did not support the assumption of genome-scale oscillations. Here, we proposed a computational model for the stochastic transitions of enhancer methylations during cell cycling. The model combines random distribution of methylation marks in DNA replication, autocatalysis of DNA methylation due to the binding of DNMT3a/b, and the collaboration between neighboring enhancers during the reconstruction of methylation marks. The proposed model can nicely explain the transition of methylation level and heterogeneous methylation distributions. Model simulations shown that the proper values of the autocatalysis is important for the heterogeneity between different cells, and increasing the collaboration between neighboring enhancers can promote the heterogeneity.  Our model suggest that methylation heterogeneity is a nature consequence of stochastic transition of DNA methylation between cell cycles and the collaboration between CpGs, however the assumption of genome-scale oscillations might not necessary for the observed heterogeneous methylation distributions.

The proposed model mainly considers the dynamics of enhancer methylation levels in a cell, and omits the biochemical reactions involved in methylation or demethylation, which may be regulated by various enzymes. In the model, we assumed a beta distribution that connects the methylation level in daughter cells with those of the mother cell. The beta distribution can take different forms based on the shape parameters that are defined by the state of mother cells. Thus, the proposed model framework mainly focus at the general effect of methylation state transition between cell cycles, while omit the detail biochemical reactions.  On the other hand, despite the complex biochemical reactions, they mainly affect the methylation levels through the based methylation/demethylation processes and the collaboration between CpGs, and hence may end up to the function $\phi$ in the model. Hence, the proposed model provides a general framework to sum up different level biochemical reactions model for DNA methylation.

\section*{Acknowledgments}
This work is supported by the National Natural Science Foundation of China (91730301, 11831015, and 11372017).

\section*{Disclosure of Potential Conflicts of Interest}
No potential conflicts of interest were disclosed.


\end{spacing}

\end{document}